# Proposal for a New Michelson-Morley Experiment Using a Single Whispering Spherical Mode Resonator

Michael Edmund Tobar, John Gideon Hartnett and James David Anstie

*Abstract*—A new Michelson-Morley experiment is proposed by measuring the beat frequency of two near degenerate modes with orthogonal propagation in a single spherical resonator. The unique properties of the experiment allow: 1. Substantial common mode rejection of some noise sources: 2. Simple calculation of the signal if Special Relativity is violated. We show that optimum filtering may be used to increase the signal to noise ratio, and to search for a preferred direction of the speed of light. Using this technique we show that a sensitivity limit of order $7 \times 10^{-19}$ is possible by integrating data over one month. We propose methods to veto systematic effects by correlating the output of more than one experiment.

*Index Terms*— Michelson-Morley, Frequency Standards, Cavity Resonators, Special Relativity.

## INTRODUCTION

The foundations of Special Relativity (SR) are based on the concept of Local Lorentz Invariance (LLI)[1]. LLI is tested by local non-gravitational experiments, and so far all tests of LLI have confirmed the validity of SR to very high precision. One of the best tools for precise measurements is the frequency standard, or "modern clock". A modern clock derives time from a well-defined frequency of an electromagnetic oscillator[2], and by searching for frequency shifts under the appropriate conditions, LLI may be tested.

Standard LLI tests include Michelson-Morley (MM)[3-5], Kennedy-Thorndike (KT)[5,6] and one-way speed of light experiments[7,8]. This work is concerned with MM experiments, which test for directional variation in the speed of light, c. This is achieved by generating two frequencies that depend on the dimension of a resonator (electronic standard), and have orthogonal propagation. When the experiment is rotated, the difference frequency is monitored to test for violations. A standard MM experiment uses two separate orthogonally orientated resonators. Modern experiments lock a microwave or laser oscillation to a high-Q resonator, with known mode structure and propagation[3-5]. The resonator is rotated and compared to a $90^0$ orientated resonator-oscillator.

We propose an alternative method using near degenerate "Whispering Spherical" (WS) modes in a single spherical microwave resonator. In particular, we show that a new class of modes exist, which we name "Whispering Longitude" (WL). This mode propagates in an orthogonal direction to the well-known Whispering Gallery (WG) mode. Measuring the beat frequency as the resonator is rotated constitutes a MM experiment. Because the modes exist in the same cavity, significantly more common mode rejection of environmental influences can be expected.

The response to violations in Special Relativity (SR) has been calculated, which includes the way the experiment is rotated and the precession of the sidereal frame. The properties of the spherical system have allowed the determination of the direction of rotation in the laboratory frame to achieve maximum sensitivity to violations in SR. For this orientation an expression is derived, which shows how the phase and amplitude of the signal varies as the light direction with respect to the preferred frame changes. Because the form of the expected signal is known, it is possible to implement optimal filtering techniques to post-process the data. Long periods (i.e. 1 to 2 years) of integration of the signal could be accommodated and significantly improve the signal to noise ratio. By changing the filter parameters a search for the direction of the laboratory frame with respect to the supposed preferred frame may also be undertaken. Furthermore, this technique may be adapted to search for more general test theories of SR than previously proposed. We show that a factor of 10 improvement over previous best MM experiments is possible with only one minute of integration, by utilizing state-of-the-art electronic resonator frequency standards.

Experimental and calculated mode properties for WS modes are presented. We show that certain WS mode families have nearly an order of magnitude greater geometric factor than the modes selected for the SUperconducting Microwave Oscillator (SUMO) project[4]. Thus, the utilization of these modes would also allow the added benefit of an order of magnitude increase of Q-factor for these types of experiments.

## ROBERTSON-MANSOURI-SEXL FRAMEWORK

Useful frameworks to test for violations of LLI have been developed by Robertson[9] and Mansouri and Sexl[10]. This framework describes a more general transformation between a moving frame $S$ of velocity $\underline{v}$ with respect to a preferred frame $\Sigma$ (see fig. 1). The speed of light, $c$, is assumed to be constant in $\Sigma$ and in general transforms to a non-constant value, $c_S(\theta,v)$, in S.

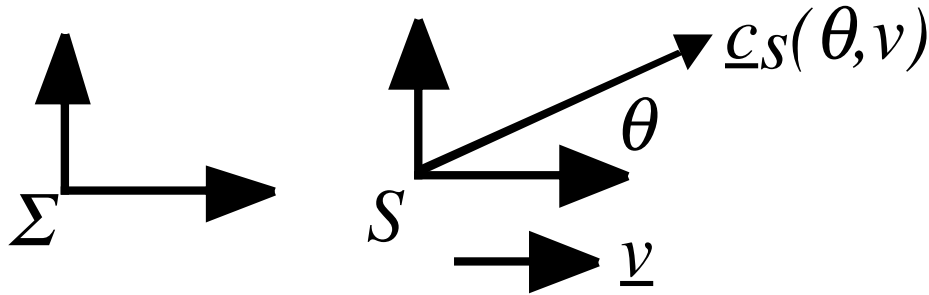

Fig. 1. Schematic of the moving frame, S, with respect to the preferred frame $\Sigma$.

In general $c_S(\theta,v)$ depends on the angle ($\theta$) of light propagation with respect to the direction ($\underline{v}$) and magnitude (v) of $\underline{v}$. The fractional difference between $c$ and $c_S(\theta,v)$ according to this transform is given by the Robertson-Mansouri-Sexl (RMS) equation below;

$$\chi(\theta,v) = \frac{c_S(\theta,v) - c}{c} = \chi_{MM}(v)\operatorname{Sin}^2\theta + \chi_{KT}(v)$$
$$\chi_{MM}(v) = \left(\frac{1}{2} - \beta + \delta\right)\left(\frac{v}{c}\right)^2, \text{ and } \chi_{KT}(v) = (\beta - \alpha - 1)\left(\frac{v}{c}\right)^2 \quad (1)$$

Here, $\alpha$ is the time dilation parameter, $\beta$ the length contraction parameter parallel to $\underline{v}$,

and $\delta$ the length contraction parameter perpendicular to $\underline{v}$. Special Relativity is a subset of this transform with $\alpha = -1/2$, $\beta = 1/2$ and $\delta = 0$, in this case $\chi(\theta, v) = 0$. Current knowledge of these parameters comes from modern MM and KT experiments. The MM experiment tests the first term in (1) while the KT experiment tests the second term. Because of the $Sin^2(\theta)$ dependence, length contractions will occur at twice the frequency of the rotation. Also, it is usual to assume that $\underline{v}$ is the Earth's velocity with respect to the cosmic microwave background[11].

NATURE OF MODES IN A SPHERICAL RESONATOR

Separation of variables of Maxwell's Equations in spherical co-ordinates ($\rho, \theta, \phi$) leads to the solution of the radial field components[12] $E_\rho$ and $H_\rho$. The mode solutions can be classified as Transverse Magnetic ($TM_{npm}$) or Electric ($TE_{npm}$) with respect to the radial direction, with the respective radial components of TM (2) and TE (3) modes given by;

$$E_\rho = n(n+1)\frac{\sqrt{k\rho}}{\rho^2} J_{n+\frac{1}{2}}(k\rho) P_n^m(Cos\theta) \begin{matrix} Cos(m\phi) \\ Sin(m\phi) \end{matrix} \quad H_\rho = 0 \qquad (2)$$

$$H_\rho = n(n+1)\frac{\sqrt{k\rho}}{\rho^2} J_{n+\frac{1}{2}}(k\rho) P_n^m(Cos\theta) \begin{matrix} Cos(m\phi) \\ Sin(m\phi) \end{matrix} \quad E_\rho = 0 \qquad (3)$$

Here $n$ is the order of the spherical Bessel function, $k$ is the wave number, $m$ the number of $2\pi$ variations along the azimuth and $p$ is the number of radial variations (which depends on $k$). The other field components ($H_\theta$, $H_\phi$, $E_\theta$ and $E_\phi$) may be found from Maxwell's equations. The $\theta$ and $\phi$ terms for all field components are cyclic in $2\pi$ and are hence common to all spherical resonators with or without dielectric and cavity walls. Frequencies are determined by solving the boundary value problem in $\rho$, and thus depend on the dielectric and metallic boundaries. Tangential fields are matched at these boundaries to solve the problem. In the following analysis we use the example of an empty spherical cavity resonator of 5-cm diameter and compare the modes to the SUMO $TM_{010}$ mode in a cylindrical resonator.

To create a stable frequency, a high-Q resonator is necessary. Superconducting resonators[13] with low surface resistance, *Rs*, have produced Q-factors as high as $10^{11}$. The Q-factor is inversely proportional to *Rs*, the proportionality constant is known as the geometric factor (or G-factor), *G*, and is dependent on the mode geometry.

$$Q = G/Rs \qquad (4)$$

The G-factor may be calculated from the ratio of the tangential H-field at the conductor surface, to the H-field in the volume of the resonator, and is given by;

$$G = \frac{\omega \iiint_V \mu_0 \overline{H}^* \cdot \overline{H} \, dV}{\oiint_S \overline{H}_t^* \cdot \overline{H}_t \, ds} \qquad (5)$$

Other important related parameters are the electric ($Pe_i$) and magnetic ($Pm_i$) energy filling factors given in (6).

$$Pe_i = \frac{\iiint_V \varepsilon |E_i|^2 \, dV}{\iiint_V \varepsilon \underline{E}^* \cdot \underline{E} \, dV} \quad Pm_i = \frac{\iiint_V \mu |H_i|^2 \, dV}{\iiint_V \mu \underline{H}^* \cdot \underline{H} \, dV} \qquad (6)$$

They represent the fraction of field energy to the total field energy, where i = ρ, θ or φ. Fig. 2 compares the G-factor of the SUMO resonator to modes in a spherical resonator and table 1 shows the filling factors for the $TE_{61m}$ WS mode family. WS TE modes have most of the H-field in the $H_\rho$ component and little in the tangential components, $H_\theta$ and $H_\phi$. Thus, they have higher G-factor than the TM modes, and a much larger G-factor than the SUMO resonator. For example, the $TE_{61m}$ mode has a geometric factor 8 time greater than the SUMO resonator. Fig. 2 also reveals that the G-factor versus frequency for WS TE modes is proportional to frequency. It turns out this is true for all TE modes in the sphere, independent of the mode numbers, n, p and m.

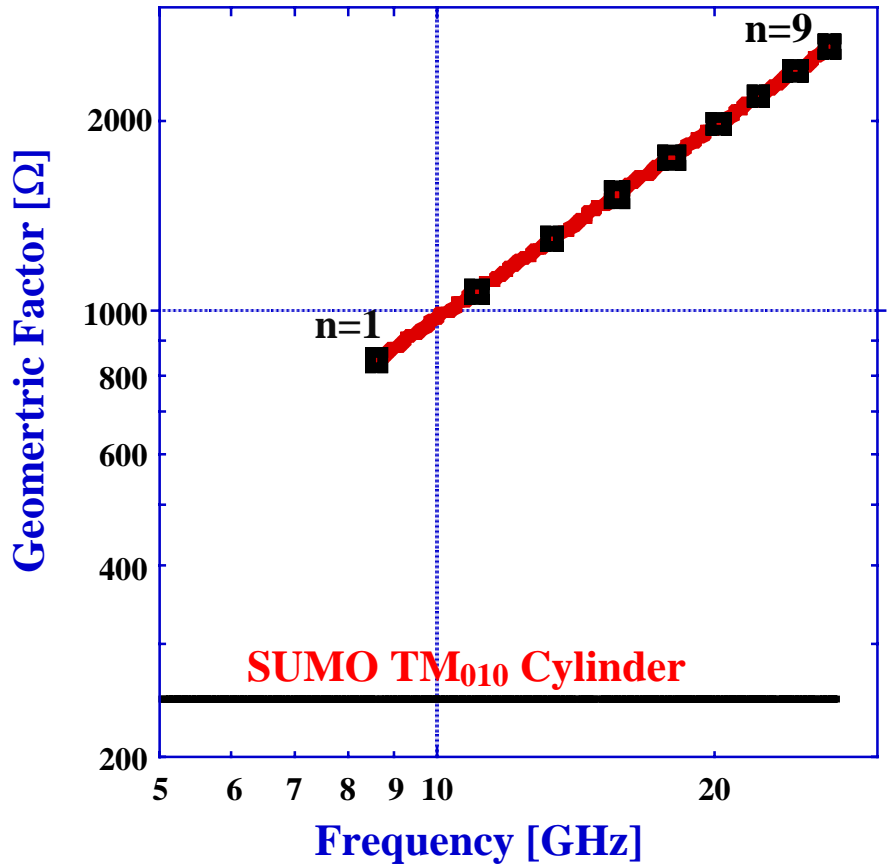

Fig. 2. Calculated Geometric factor versus frequency for WS $TE_{n,1,m}$ modes in a sphere of diameter 5-cm, compared to the $TM_{010}$ mode in a cylinder of the same aspect ratio of SUMO. The diameter of the SUMO resonator fixes the frequency at 8.6 GHz. Each point represents a degenerate mode family given by (2) and (3). If the sphere is perfect each member, $m = 0$ to $n$ has the same frequency and Q-factor.

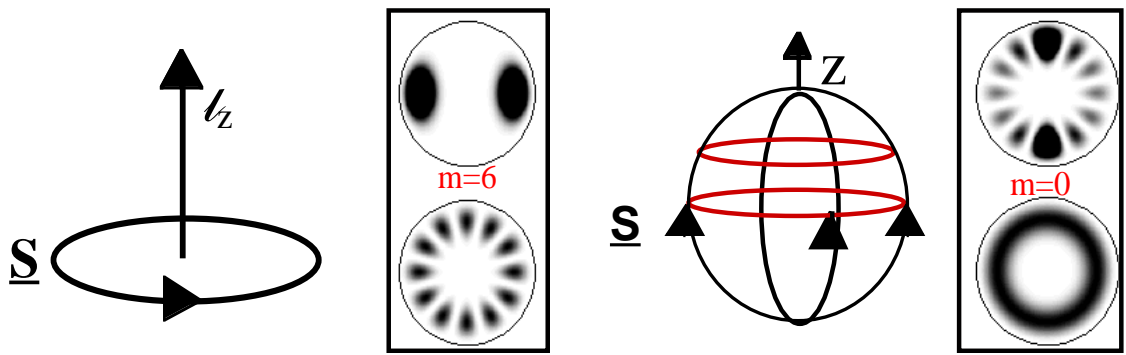

Fig. 3. Left, $TE_{616}$ WG mode: Right, $TE_{610}$ WL mode. The Poynting vector, $\underline{S}$, shows the direction of propagation on the sphere surface. The WG mode propagates like a ray around the equator with angular momentum in the z direction. The WL mode propagates as wave fronts along the longitude, which converges at the intersection with the z-axis. This results in a non-degenerate ($m = 0$) 'self' standing wave with maximum intensity along the z-axis. The density plots show $|H_\rho|^2$, above is in the $z$-$y$ plane and below is in the $x$-$y$ plane.

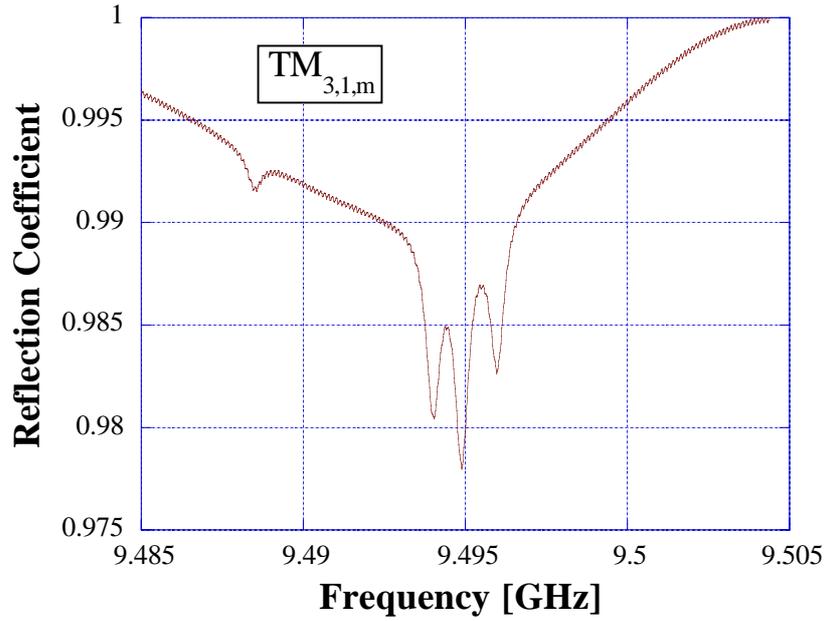

Fig. 4. Measure reflection coefficient for the $TM_{31m}$ near-degenerate mode family.

TABLE I

ELECTRIC AND MAGNETIC FILLING FACTORS FOR THE $TE_{61M}$ MODES

| m | $Pe_\theta$ | $Pe_\phi$ | $Pm_\rho$ | $Pm_\theta$ | $Pm_\phi$ |
|---|---|---|---|---|---|
| 6 | 0.926 | 0.074 | 0.992 | 0.001 | 0.008 |
| 5 | 0.749 | 0.251 | 0.992 | 0.002 | 0.006 |
| 4 | 0.556 | 0.444 | 0.993 | 0.003 | 0.004 |
| 3 | 0.361 | 0.639 | 0.993 | 0.005 | 0.003 |
| 2 | 0.185 | 0.815 | 0.993 | 0.006 | 0.001 |
| 1 | 0.055 | 0.945 | 0.992 | 0.007 | 0.000 |
| 0 | 0.000 | 1.000 | 0.991 | 0.009 | 0.000 |

From the filling factors (see Table I) it is straightforward to deduce that WS modes propagate predominately around the spherical surface. Since $Pm_\rho \sim 1$ and $Pe_\rho = 0$, the cross product between $\underline{E}$ and $\underline{H}$ will give a Poynting vector ($\underline{S}$) with predominately $S_\theta$ and $S_\phi$ components. It is necessary to choose modes with orthogonal propagation for a MM experiment. For a given n, a set of degenerate WS modes exist, with m = 0 to n. In particular modes with $m = 0$, propagate along the longitudes, while the modes with $m = n$ propagate along the azimuth. A schematic showing $\underline{S}$ and $H_\rho$ is shown in figure 3. The other modes ($m$ = 1 to $n$-1), propagate with a combination of directions. Experimentally we confirmed the presence of $n+1$ near degenerate modes for each n. For example, the $TM_{31m}$ mode existed with 4-fold degeneracy with frequencies of 9.48851, 9.49399, 9.49488 and 9.49597 GHz, these compare well with the calculated value of 9.4919. The frequency separation is due to imperfections of the spherical surface. The reflection coefficient for this set of modes is shown in fig. 4.

### SENSITIVITY TO TWO-WAY SPEED OF LIGHT ANISOTROPY

To determine the sensitivity of WL and WG modes to violations of SR, the average speed of light the modes sample must be calculated. First we assume the WG and WL modes share the same z-axis (this has been proven experimentally by the authors). The resonator axis (laboratory frame S) is labeled as z' and assumed to be at an angle $\theta_z$ to the preferred frame (direction of $\underline{v}$ (z) in the absolute frame $\Sigma$). Eq. (1) is an ellipsoid and must be calculated in the rotated frame. In general a speed of light ellipsoid, $c_S(\phi,\theta)$, may be represented by equation (7), where $c_x$, $c_y$ and $c_z$ are the light speed of along the Cartesian directions of the preferred frame.

$$\begin{bmatrix} x & y & z \end{bmatrix} \begin{bmatrix} 1/c_x^2 & 0 & 0 \\ 0 & 1/c_y^2 & 0 \\ 0 & 0 & 1/c_z^2 \end{bmatrix} \begin{bmatrix} x \\ y \\ z \end{bmatrix} = 1 \qquad (7)$$

This is of the form; $\underline{x}^t \underline{\underline{X}} \underline{x} = 1$. To be consistent with the RMS equation (1), we substitute the following; $c_z = c$, $c_x = c + \Delta c_x$, $c_y = c + \Delta c_y$ and set $\Delta c_x = \Delta c_y = \chi_{MM}(v)$, and convert to spherical co-ordinates using $z = c_S(\phi,\theta)Cos\theta$, $x = c_S(\phi,\theta)Sin\theta\, Cos\phi$ and $y = c_S(\phi,\theta)Sin\theta\, Sin\phi$. After these substitutions, the solution of $c_S(\phi,\theta)$ from (7) reduces to (1) (ignoring the KT term). We note that the equations presented here may be adapted to test more general test theories, such as bi-axial anisotropy when $\Delta c_x \neq \Delta c_y$.

To calculate the speed of light ellipsoid in the laboratory frame, the preferred frame must be arbitrarily rotated. A schematic 3-D plot of $c_s(\phi,\theta)/c - 1$ in the rotated frame is shown in fig. 5. This may be calculated using the following transformation.

$$\underline{x}' \underline{\underline{A}} \underline{\underline{X}} \underline{\underline{A}}^{-1} \underline{x}' = 1 \tag{8}$$

Here $\underline{\underline{A}}$ is the relevant rotation matrix and x' is the x vector in the dashed frame (depends on the angles ψ or $\gamma_1$ and $\gamma_2$). Following this approach the speed of light ellipsoid may be represented by (details of the calculation are not presented here, and may be obtained from the authors):

$$\chi_{wg}(\theta_z,\psi,v) = \frac{c_s(\theta_z,\psi,v)}{c} - 1 = \chi_{MM}(v)\left(Cos^2\psi Cos^2\theta_z + Sin^2\psi\right) \tag{9a}$$

$$\chi_{wl}(\theta_z,\gamma_1,\gamma_2,v) = \frac{c_s(\theta_z,\gamma_1,\gamma_2,v)}{c} - 1$$
$$= \chi_{MM}(v)\begin{pmatrix} Cos^2\gamma_1 Cos^2\theta_z Sin^2\gamma_2 + Sin^2\gamma_1 Sin^2\gamma_2 + \\ Sin^2\theta_z Cos^2\gamma_2 + 2Cos\gamma_1 Cos\gamma_2 Cos\theta_z Sin\gamma_2 Sin\theta_z \end{pmatrix} \tag{9b}$$

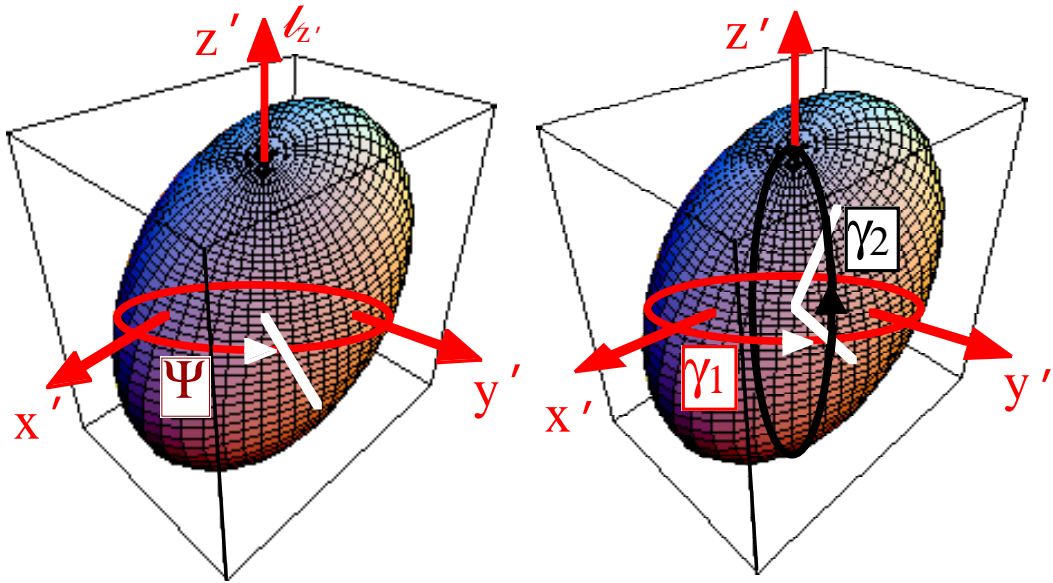

Fig. 5. Speed of light ellipsoid given by (9) rotated with respect to the WS resonator frame.

To calculate the average speed of light the WG and WL modes sense at any instant, (9) must be integrated in the resonator (dashed) frame (see fig. 5) and are given by;

$$\chi_{wg}(\theta_z,v)_{av} = \frac{1}{\pi}\int_0^\pi \chi_{wg}(\theta_z,\psi+\pi/2,v)\,d\psi = \chi_{MM}(v)\left(\frac{1+Cos^2\theta_z}{2}\right) \quad (10a)$$

$$\chi_{wl}(\theta_z,v)_{av} = \frac{1}{\pi}\int_0^\pi \left(\frac{1}{\pi}\int_0^\pi \chi_{wl}(\theta_z,\gamma_1,\gamma_2+\pi/2,v)\,d\gamma_2\right)d\gamma_1 = \chi_{MM}(v)\left(\frac{2+Sin^2\theta_z}{4}\right) \quad (10b)$$

Note that a π/2 phase shift must be added along the angle that defines the direction of propagation, this is due to the direction of light being perpendicular to the actual co-ordinate of the ellipsoid. This phase shift actual makes no difference to the result of the integral when averaged over the whole resonator. The beat frequency between the WL and WG modes will be given by the difference between (10a) and (10b), and may be calculated to be:

$$\frac{\Delta f|_{WG-WL}}{f} = \frac{1}{4}\chi_{MM}(v)\left(2Cos^2\theta_z - Sin^2\theta_z\right) \quad (11)$$

Because the modes have different propagation and share the same z-axis, (11) is only dependent on $\theta_z$. For standard MM experiments this expression depends on two angles of rotation with respect to the absolute frame rather than the position of the z'-axis, and is thus somewhat more complicated. This is because the standard experiments use two of the same modes orientated at 90°. (The author has verified this in the case of two 90° orientated WG modes and SUMO type resonators). Consequently, if we spin the WS mode resonator around its z-axis on time scales faster than the rotation of the earth, then the experiment will be insensitive to uniaxial violations in SR even if they exist, as $\theta_z$ does not change. To obtain maximum sensitivity, the resonator must be spun around the axis perpendicular to z'. Even though we have determined the way to rotate the experiment, the sensitivity will still depend on the angle, $\theta_i$ between $z$ and $z'$ in the x-y plane, as well as the direction of rotation, $\lambda$, as shown in fig. 6.

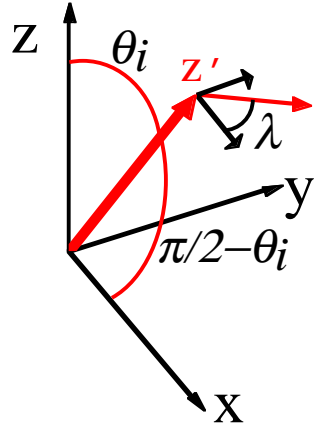

Fig. 6. In a MM experiment the z-axis of the WL and WG modes ($z'$ vector shown) is rotated with respect to the absolute frame. The rotation may be defined by two angles, $\theta_i$ and $\lambda$. Here $\theta_i$ is the angle between the z- and z'-axis in the z-x plane, and $\lambda$ defines the direction of the rotating vector with respect to the x'-axis.

The response to rotating the experiment was calculated by applying the relevant rotational co-ordinate transforms of the z-axis to the z'-axis in the following order: 1. The z-axis is rotated by $\theta_i$ around the y-axis: 2. The new x' and the y axis are rotated around z' by $\lambda$: 3. The z' axis is rotated around the y' axis at an angle equal to $\omega t$. The form of the beat frequency between the WL and WG modes can be calculated by first taking the dot product between the z and z' axis to calculate $\theta_z(t)$. Following this $\theta_z$ is substituted back into (11) to give.

$$\frac{\Delta f|_{WG-WL}}{f} = \frac{3}{8}\chi_{MM}(v)\left(1 - Sin^2\theta_i Sin^2\lambda\right)Cos[2\omega t + \vartheta(\theta_i,\lambda)]$$

$$\vartheta(\theta_i,\lambda) = Cos^{-1}\left[\frac{Cos^2\theta_i - Cos^2\lambda\; Sin^2\theta_i}{Cos^2\theta_i + Cos^2\lambda\; Sin^2\theta_i}\right]$$

(12)

Thus, violations in SR will manifest as a $2\omega$ variation of the beat frequency. Also, the amplitude and phase will be modulated due to the precession of the earth's sidereal frame (due to changing $\theta_i$ and $\lambda$), which must be taken into account.

The known form of the expected signal may be used to our advantage in the post-processing of the data. Standard optimal filter techniques may be used to search for a signal of known characteristic amongst the noise. In general there are three unknown parameters we must vary during our search for violations in SR, i.e. v, θ$_i$ and λ. These three parameters will change due to the rotation of the sidereal frame. The first term in (10), $\chi_{MM}(v)$, will cause a relatively small amplitude modulation at the sidereal frequency in a similar way to a KT experiment, while the θ$_i$ and λ dependent terms will cause large amplitude and phase modulation at twice the sidereal frequency. Also, it is feasible that this technique may be extended to search for more complicated violations in SR by calculating new filter templates, (i.e. the response to biaxial symmetric violations has been calculated by the authors). Biaxial violations would occur if the supposed length contraction parameter perpendicular to v, δ, had a preferred direction itself (i.e. it is a function of the azimuth angle (φ) in the plane perpendicular to the motion).

Optimal filter templates can be constructed from the expected signal and known noise in the system. Different templates may be applied to the same data to undergo a search for unique values of v, θ$_i$ and λ that produce a statistically significant signal above all other values. Also a template for biaxial violations could be constructed. If there does exist violations in SR, it is important that the phase and amplitude variations assumed are correct, otherwise the effective integration (or filtering) of the signal may not increase the signal as expected and hence the upper limits to violations in SR cited may not be correct. This is especially important as the time for integrating the signal is increased.

Estimation of the sensitivity is achieved by calculating the average response from (12). Given that the average of sine squared over one cycle is 1/2. Then the average signal response is:

$$\left\langle \frac{\Delta f}{f} \right\rangle = \frac{9}{32} \chi_{MM}(v) Cos[2\omega t + \vartheta(\theta_i, \lambda)] \qquad (11)$$

The spectral density of frequency fluctuations is typically, $\sqrt{S_y} \sim 10^{-16}$ Hz/√Hz at 1 Hz

Fourier frequency, for a state-of-the-art oscillators. Thus, fractional frequency changes of the order, $10^{-16}$ may be measured with a SNR=1 on time scales of order 1 s. In addition, it is well known that the SNR of a cosine signal amongst stationary noise, is increased by the square root of the observation time, $t_{obs}$. Thus, the sensitivity for long integration times, ($t_{obs} \gg 1$ day) may be estimated from:

$$\chi_{MM}\big|_{SNR=1} = \frac{32}{9} \sqrt{\frac{S_y(f)}{t_{obs}}} \qquad (12)$$

From (12) we conclude that a measurement may be made on average (depending on the orientation with respect to the absolute frame) with $10^{-16}$ sensitivity with only one minute of data. This limit is more than one order of magnitude better than the previous best measurement[3]. For one month of integration, the limit is calculated to be $7 \times 10^{-19}$. It should be noted that this estimate depends on the noise spectral density remaining stationary over the integration. Non-stationary effects could mimic the signal to some degree and cause systematic errors. This has shown to be true when searching for known gravitational wave signals amongst noise. The best way to veto these "fake" signals is to correlate the output of two detectors. Some simple ways this may be done is discussed in the following paragraph.

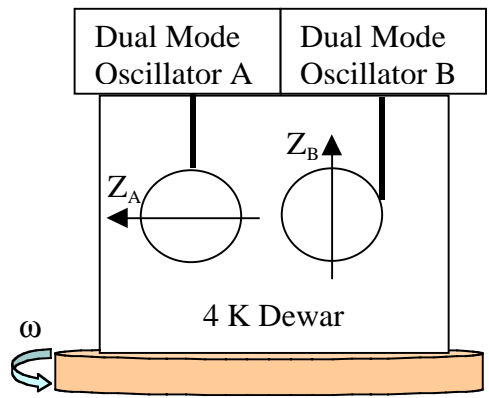

Fig. 7. Schematic of the proposed MM experiment. This will comprise two spherical high-Q resonators cryogenically cooled to 4K and orientated in orthogonal directions as shown. The experiment will be placed on a rotation table and rotated at an angular frequency of $\omega$ rads/s. Oscillator A will have maximum sensitivity to violations in SR at a frequency of $2\omega$, while oscillator B will have zero sensitivity. Oscillator B will provide data to check for systematic errors in the statistics of the data analysis, as well as a measurement to test a new test theory based on biaxial violations of SR.

We propose to perform a MM experiment with spherical resonators made from superconducting niobium or sapphire monocrystal. Implementing resonators based on these materials has produced highly stable microwave oscillators, which exhibit frequency instabilities of order 1 part in $10^{16}$ at short time intervals of order 1 to 100 seconds[13,14]. The resonator will be operated in dual mode configuration that will simultaneously excite the WL and WG modes. The dual-mode technique has been developed independently as a new method to attain frequency stability in an anisotropic system [15], and will be straightforward to adapt to this situation. Rejection of systematic errors could be expected by placing two spherical resonators in the same cryogenic dewar, with one orientated to attain maximum sensitivity (system A) and the other to obtain zero (system B) as shown in fig. 7. This would enable testing the statistical significance of the integrated signal as a function of different values of $\theta_i$ and $\lambda$. If a violation in SR is observed one would expect a stronger signal from the A system, with System B used as a check for systematic errors and statistical significance. Another check on systematic errors could be achieved by operating two or more simultaneous MM experiments. Signals from all the systems may be compared by correlating the output for differing values of $\theta_i$ and $\lambda$ according to (10). One would expect to measure the same effect in all systems if Special Relativity was violated.

## CONCLUSION

If a spherical sapphire or superconducting oscillator is realized with state of the art frequency instability of $10^{-16}$, then violations in SR may be measured on the order of $7 \times 10^{-19}$, by integrating a signal over one month. Also, the direction of a preferred axis may be tested for, by simply post-processing the data for characteristic amplitude and phase modulation. The quoted sensitivity is only a limit due to stationary noise. Non-stationary noise could cause systematic errors that mimic signals. We have proposed correlating the output of more than one system to veto these effects.

To be published in Phys. Let A. 2002.

ACKNOWLEDGEMENTS

This work was supported by the Australian Research Council. Also, many thanks to Dr. Frank van Kann for many useful discussions.

REFERENCES

1  C. M. Will, *Theory and Experiment in Gravitational Physics* (Cambridge University Press, Cambridge, 1993).

2  C. Salomon, N. Dimarcq, *et al.*, C. R. Acad. Sci. Paris **IV**, 1-17, 2001.

3  A. Brillet and J. L. Hall, Phys. Rev. Lett. **42**, 549-552, 1979.

4  S. Buchman, *et al.*, Adv. Space Res. **25**, 1251-1254, 2000.

5  C. Lammerzahl, , *et al.*, Class. Quantum Grav. **18**, 2499-2508, 2001.

6  D. Hils and J. L. Hall, Phys. Rev. Lett. **64**, 1697-1700, 1990.

7  T. P. Krisher, L. Maleki, *et al.*, Phys. Rev. D **42**, 731-734, 1990.

8  P. Wolf and G. Petit, Phys. Rev. A **56**, 4405-4409, 1997.

9  H. P. Robertson, Rev. Mod. Phys. **21**, 378, 1949.

10  R. Mansouri and R. U. Sexl, Gen. Relativ. Gravit. **8**, 497-514, 1977.

11  G. F. Smoot, M. V. Gorenstein, R. Muller, Phys. Rev. Lett. **39**, 898, 1977.

12  J. A. Stratton, *Electromagnetic Theory* (McGraw-Hill, New York, 1941).

13  S. R. Stein and J. P. Turneaure, Proc. of IEEE **36**, 1245 (1975).

14  S. Chang, A. Mann, A. Luiten, Electron. Lett. **36**, 480-481, 2000.

15  M. E. Tobar,  *et al.*, in Proc. IEEE IFCS, 2001.